# Reversible and irreversible changes in physical and mechanical properties of biocomposites during hydrothermal aging

Arnaud Regazzi, Stéphane Corn*, Patrick Ienny, Jean-Charles Bénézet, Anne Bergeret

École des Mines d'Alès, C2MA, 6 avenue de Clavières, F-30319 ALES CEDEX, France

*Corresponding author. Tel.: +33 4 66 78 56 29; fax: +33 4 66 78 53 65, e-mail address: Stephane.Corn@mines-ales.fr*

## ABSTRACT

The use of biocomposites on a daily basis for industrial products brings to light the influence of environmental factors on the evolution of mechanical properties. This aging has a major influence in the lifetime of any product based on such materials. Among biobased composites, poly(lactic acid) reinforced with plant fibers are known to be sensitive to hydrothermal aging due to the intrinsic nature of their components. Although some papers studied the influence of temperature and water absorption on such materials, so far the difference between reversible and irreversible effects of aging has been hardly studied. This distinction is the purpose of this study. Poly(lactic acid) samples reinforced with various content of flax fibers (0%, 10%, and 30%) were immersed in water at different temperature (20, 35, and 50°C) up to 51 days. The physical and chemical phenomena responsible for the changes in the mechanical properties of biocomposites were evaluated. It was observed that these changes were mainly reversible. However, irreversible effects of aging turned out to increase drastically with the amount of fiber and the aging temperature. While hydrolysis drastically deteriorated PLA for longer aging times, the lifetime of biocomposites was significantly extended (+ 230% at 50°C from 0% to 10% in fiber content) by the presence of fibers which postponed the failure of PLA.

## 1. INTRODUCTION

The growing demand for sustainable policies from public opinion drives industry toward the development of bio-based materials. In the field of thermoplastic composites, poly(lactic acid) reinforced with plant fibers meets a growing success for industrial application because of their good specific mechanical properties, availability, and similar processing properties than traditional polymers (Bocz et al., 2014). A limited number of products can already be found on the market targeting various fields of application such as automotive, mobile phones, or plant pots (Graupner et al., 2009; Summerscales and Grove, 2014). Such products are fully biodegradable and offer the advantage to be either recycled, composted or incinerated at their end of life (Le Moigne et al., 2014). However the intrinsic nature of these materials results in a high sensitivity to the presence of water and temperature (Le Duigou et al., 2009). Indeed several papers highlight the specific problems encountered in the presence of a hydrothermal environment (Islam et al., 2010) resulting in a decrease of mechanical properties. Though they do not rule out the idea of using such materials for outdoor applications (Cheung et al., 2009; Dittenber and GangaRao, 2012). As a result, the success of the commercial use of biobased composites lies in the control of the evolution of their mechanical properties in real-life conditions. This evolution of properties (physical and mechanical), which are the effects of ageing, can be either temporary (reversible) or permanent (irreversible). In order to control aging, the first step consists in identifying the aging factors susceptible to result in a change of these properties. Among them, bacterial degradation, UV radiations and mechanical loadings are not negligible but temperature and humidity are the major causes for short-term aging (Kumar et al., 2010; van den Oever et al., 2010). The processes induced by the aging factors and their combination are well identified since they have been studied for years for usual polymers. However, their extension to biobased polymers requires additional research efforts because of the numerous processes at stake, their interactions, and the heterogeneity of both materials and aging processes. In addition to these difficulties, aging effects may be reversible, like plasticizing for instance, or, in most cases, irreversible (Mercier et al., 2008; Pochiraju et al., 2012; Weitsman, 2012). This distinction is essential to predict the evolution of properties, especially when aging factors are expected to evolve during the materials lifetime. Indeed, for any outdoor application, aging is cycled: day/night, sun/rain, seasons, indoor storage/outdoor use. In order to properly assess the sensitivity of materials to aging in real-life conditions within a time period of several years, a complete understanding of the impact of both reversible and irreversible effects of aging is necessary. The purpose of this study is to assess the consequences of both the reversible and irreversible effects of a hydrothermal aging on the properties of a plant fiber reinforced poly(lactic acid) in order to offer new perspectives for predicting the lifetime of natural fibers reinforced composites.

## 2. MATERIALS AND METHODS

### 2.1. Methodology

Poly(lactic acid) / short flax fiber biocomposites have been selected for the present study. As described in Figure 1, the analysis was decomposed into two steps:

1. Firstly the impact of hydrothermal aging was assessed on composites with various fiber contents. In this case, reversible and irreversible effects of aging occurred simultaneously, and only their combined effects were evaluated.

2. In a second step, materials were brought to the same hydrothermal conditions as before aging by removing water by desiccation. This procedure allowed the assessment of potential damage, and thus the distinction between reversible and irreversible effects of aging.

It is worth to mention that, in this study, the properties of the materials after an initial desiccation effected prior to aging constitutes the reference for the analysis of their changes. The total change of a property due to aging corresponds to the variation between its value after immersion and its reference value. Accordingly, the irreversible change of a property is defined by its variation at final desiccation. Then, the reversible part of a property is assessed from the difference between its total change and its irreversible change.

## 2.2. Materials

### 2.2.1. Poly(lactic acid) matrix

PLA Ingeo™ 7000D resin was obtained from NatureWorks® LLC (Blair, NE, USA). These granules were designed for injection stretch blow molded applications. This grade of PLA had a density of 1.24 g/cm³, a glass transition temperature between 55 and 60°C and a melting temperature between 155 and 165°C (NatureWorks, 2005).

### 2.2.2. Flax fibers

The short flax fibers (*Linum usitatissimum*) FIBRA-S®6A used for this study were provided by Fibres Recherches Développement® (Troyes, France). According to the technical data sheet (FRD, 2011), fibers (bundles) were 6 mm long with a diameter of 260 ±150 μm and their density was between 1.4 and 1.5 g/cm³. Concerning quasi-static mechanical properties, Young's modulus of bundles was 36 ± 13 GPa, maximum stress was 750 ± 490 MPa and strain at break was 3.0 ± 1.9 %.

## 2.3. Experimental techniques

### 2.3.1. Processing conditions

Various fiber weight contents were used: 0% (neat PLA) hereafter named PLA, 10% hereafter named PLA-F10, and 30% hereafter named PLA-F30. Polylactic acid granules and flax fibers were dried at 80°C for at least 24h and under vacuum at 120°C for 4h respectively. Composite granules were obtained with a corotative twin-screw extruder (Clextral BC21, screw length = 900 mm; temperature profile along the screw and at the die = 180°C). After a second drying step under vacuum at 80°C during 24h, compounded granules were molded with an injection molding machine (Krauss Maffei KM50-180CX) into dog-bone samples according to the standard ISO 527-2 1BA. The temperature profile was increasing up to 200°C and the mold was kept at 25°C. After processing, samples were stored at room temperature and 2%rh (relative humidity) before characterization or aging. This equilibrium state constitutes the reference for evaluating the effects of aging on the materials.

### 2.3.2. Aging experiments

Three series of isothermal water immersion experiments were conducted for each material (PLA, PLA-F10 and PLA-F30) at 20°C, 35°C, and 50°C. For each temperature and material, 10 samples were immersed in water, slightly wiped, characterized then re-immersed, up to 51 days. Afterward, samples were desiccated at room temperature and 2%rh until their mass reached an equilibrium. This desiccation was supposed not to imply other mechanisms than those previously caused by hydrothermal aging. This equilibrium state constitutes the final condition for evaluating the irreversible effects of aging.

### 2.3.3. Weight, volume, and density measurements

Samples mass was measured using a weighing scale Mettler-Toledo AT200 with a precision of 0.1 mg. Thickness and width were determined with a 1 μm-accurate palmer and length with a 10 μm-accurate caliper. A mean value performed on each sample was determined. The volume of the sample was approximately evaluated on the basis of dimensional measurements. As a result, the evaluation of density was based on water uptake and volume gain measurements. The reproducibility was evaluated on 10 measurements.

### 2.3.4. Scanning electron microscope observations

The morphology of the samples after several aging durations was analyzed using an environmental scanning electron microscope (ESEM) FEI™ Quanta 200 FEG. They were freeze-fractured in the middle part, and the surface was carbon-coated with Balzers CED030 sputter-coating device.

### 2.3.5. Size exclusion chromatography

The molecular mass of PLA was evaluated by size exclusion chromatography (SEC) with Optilab® rEX™ of Wyatt Technology (CIRAD, UMR 1208 (IATE), Université de Montpellier, France). 90 mg of each aged material was diluted in tetrahydrofuran stabilized with butylated hydroxytoluene, and then kept at 30°C during 40h in a water bath. After a 45µm-filtration, each solution was injected in the column for measurement. The reproducibility was evaluated on 3 samples.

### 2.3.6. Differential scanning calorimetry

A Perkin-Elmer® Diamond DSC was used in order to assess the crystallinity of PLA. Samples mass ranged from 5 to 10 mg and ramp temperature was set at 10°C/min. The enthalpy of a 100% crystalline PLA was assumed to be 93.7 J/g in order to determine crystallinity from thermograms (Garlotta, 2001).

### 2.3.7. Vibration analysis

From a general point of view, vibration techniques aim to study the dynamic behavior of a structure using its fundamental natural mode of vibration (Corn et al., 2012; Hwang and Chang, 2000). In the present study, this technique was expected to evaluate the intrinsic viscoelastic properties of the materials and more particularly their dynamic elastic modulus (storage modulus) and damping coefficient (Adams, 1996).

Samples were set in cantilever position (one end clamped, the other end free) with a free length of 60 ± 0.2 mm. The free end was stimulated by a short impulse leading the sample to vibrate. The displacement during vibration was recorded by a laser sensor (SunX HL-C203F). The time response was then converted to a frequency response by means of a Fast Fourier Transform (FFT) (Gibson, 2000). The eigenfrequency and the damping factor of the first fundamental natural mode of vibration were determined using an interpolation of the Nyquist plot of the frequency response function. This mode corresponded to the first flexural mode. As no simple analytical solution existed for such "dog-bone"-shaped samples, a finite element analysis was conducted in order to extract the dynamic elastic modulus from the eigenfrequency.

The main advantages of this vibration technique are its non-destructive nature and its easy implementation, so that it was well suited for monitoring materials properties during aging. However, it requires the knowledge of the density and the dimensions of the sample which varied during aging. All vibration tests were carried out at room temperature whatever the sample aging conditions. The reproducibility was validated on 5 samples.

2.3.8. Uniaxial tensile tests

Ultimate tensile properties were assessed with a testing machine Zwick B Z010 according the ISO 527 standard. Samples were loaded at 5 mm/min at room temperature. The reproducibility was evaluated on 3 samples.

3. RESULTS

3.1. Physical properties during aging

Figure 2 shows experimental results for water uptake of the three materials at 20°C, 35°C and 50°C. It can be observed that both water content and diffusion kinetics were enhanced by both temperature and fiber content. At 20°C and 35°C, PLA seemed to obey to a Fick diffusion since water uptake reached an equilibrium for long immersion times (Crank, 1975). Other sorption curves (except PLA-F30 at 50°C) displayed a gradual increase at long times. This diffusion is also called "pseudo-Fick" diffusion (Weitsman, 2000) since only the beginning of the sorption follows Fick's model. However at 50°C, PLA-F30 exhibited a weight loss beyond 144 h (or 12 √h).

The volume and density variations were also assessed during these experiments. Density is plotted as a function of immersion time in Figure 3. Since the density of flax fibers was higher than the density of PLA, the initial density (unaged samples) of composites increased with the fiber content – from 1.238±0.005 g/cm³ for PLA to 1.254±0.005 g/cm³ for PLA-F10 and 1.285±0.005 g/cm³ for PLA-F30. Density turned out to decrease with immersion time with or without the presence of fibers. However this decrease was amplified by both temperature and fiber content.

### 3.2. Mechanical properties during aging

Figure 4 shows the evolution of the dynamic elastic modulus (DEM) as a function of immersion time for the three materials during immersion in water at 20, 35, and 50°C. Before aging, an increase in the initial DEM with the fiber content was observed: +24% and +95% for PLA-F10 and PLA-F30 respectively, compared to PLA. These results are in accordance with literature where mechanical properties were assessed with quasi-static measurements (Bax and Müssig, 2008; Le Duigou et al., 2008; Oksman, 2003; van den Oever et al., 2010).

During aging, the DEM of all samples decreased with immersion time. The amount of the decrease increased with aging temperature and fiber content.

- For aging at 20°C, PLA DEM remained rather constant for the duration of our tests. However the DEM of PLA-F10 and PLA-F30 decreased and reached equilibrium at long immersion times (> 400h or 20 √h). Nevertheless, at the end of the experiment, composites remained stiffer than unaged PLA.
- For aging at 35°C, the decrease in DEM was more significant, even for PLA. The DEM of composites was approximately the same as PLA at the end of the experiment.
- For aging at 50°C, the assessment of mechanical properties of PLA and PLA-F10 became impossible respectively after 144h and 750h because of a very high brittleness. But PLA-F30 was strong enough for characterization up to the end of the experiment. At long immersion times, it exhibited a DEM much lower than unaged PLA.

The evolution of ultimate tensile properties during immersion was evaluated as well (cf. Table 1). Before aging, the ultimate tensile strength of PLA, PLA-F10, and PLA-F30 was 67 ± 1, 62 ± 2, and 66 ± 4 MPa, respectively. During immersion, ultimate tensile strength decreased for all materials but more significantly as temperature and fiber content increased. Characterization tests could not be performed at 50°C due to the extreme ductility of the samples. Prior to aging, the strain at break of PLA, PLA-F10, and PLA-F30 was 5.5 ± 0.3, 3.1 ± 0.2, and 1.6 ± 0.2 %, respectively. During immersion, ductility was significantly improved regardless of the fiber content. To sum up, results showed that, during aging, strain at break increased whereas ultimate tensile strength and dynamic elastic modulus decreased, which was a characteristic of softening materials.

Figure 5 presents the evolution of damping of the first fundamental natural mode of vibration for the three materials at each aging temperature. Damping is controlled by dissipative factors such as friction at micro- and meso-scale (Meyers and Chawla, 2009).

Before aging, it appeared that energy dissipation slightly increased with fiber content. During aging, the evolution of damping with aging temperature and fiber content was similar to the evolution of water uptake (cf. Figure 2).

### 3.3. Reversibility of physical properties

All the results presented so far were the combination of reversible effects (e.g. plasticizing) and irreversible effects (e.g. damage). The measurement of the properties after desiccation suppressed the reversible effects and enabled to quantify the irreversible effects of aging. Reversibility/irreversibility was only evaluated over the first 144h of immersion because results presented previously showed that transient phenomena were mostly occurring in this duration. Another reason was the very high brittleness of some samples beyond this time which rendered characterization impossible.

Results for 144 hours of immersion are gathered in Table 2. The weight of samples after water removal was approximately the same as their initial weight whatever the fiber content and the aging temperature. The volume and the density were completely recovered after desiccation for PLA immersed at 20°C and 35°C as well as PLA-F10 at 20°C. In all other cases, the recovery was only partial and was less and less significant for higher temperatures and larger fiber contents. This partial recovery indicated the presence of damage.

### 3.4. Reversibility of mechanical properties

The evolution of the DEM for different immersion times after desiccation is presented in Figure 6. The changes were different for pure PLA and PLA/flax composites. For PLA, a slight increase in the DEM was observed as a function of aging time. As shown in Table 3, this result was more significant at higher temperatures (1.6%, 3.0% and 5.2% at 20°C, 35°C and 50°C respectively for 144h of aging). The ultimate tensile stress exhibited a similar increase for aging at 50°C (cf. Table 1). On the contrary, the strain at break drastically decreased.

For composites, desiccation only allowed partial recovering of the DEM (cf. Figure 6). The DEM dropped with aging temperature (-9.6%, -28.4% and -38.6% at 20°C, 35°C and 50°C respectively for PLA-F30 after 144h of aging) and with fiber content (-13% and -39% for PLA-F10 and PLA-F30 respectively after 144h of aging at 50°C). The ultimate tensile strength also dropped due to irreversible effects of aging and in a similar manner as DEM according to results in Table 1. Also, the increase in strain at break observed during aging turned out to be partially irreversible, especially when fiber content was significant. Except for PLA at 50°C, the irreversible effects of aging led to an improvement of the ductility. As shown in Figure 7, the reversibility of damping was fully complete whatever the fiber content and the aging temperature.

## 4. DISCUSSION

### 4.1. Reversible and irreversible mechanisms

Results showed that the strong sensitivity of PLA/flax composites to temperature and water was responsible for important changes in physical and mechanical behavior.

First of all, an increase in water content and temperature both led to an increase in polymer chain mobility (Weitsman, 2000). This mobility facilitated the penetration of water into the polymer matrix. As a result, it enhanced both water storage capacity and diffusion rate (Badia et al., 2012). Besides, in composites, these quantities were even further increased by the high affinity of lignocellulosic fibers to water (Ndazi, 2011). In the case of flax, the water content can reach up to 50% for a loose fiber (Müssig, 2010) which is much more than PLA. As a result, both temperature and fiber content led to the increase of water content and diffusion kinetics. However the weight loss of PLA-F30 at 50°C was the expression of non-reversible damage (Weitsman, 2000).

The absorption of water resulted in the samples swelling because of the physical presence of water molecules causing the deviation of chains. As water is less dense than PLA and flax fibers (cf. 2.2), the overall density tended to decrease as water penetrated the materials. Since flax fibers absorb more water than PLA, their swelling was more significant (up to 30% for a loose fiber) (Müssig, 2010). As a result, the decrease in density was more significant as well. Similarly the increase in water uptake with temperature led to a more significant swelling and a more significant decrease in density.

As shown by characterization tests after desiccation, the presence of water induced modifications of the physical properties, which remained after complete desiccation. Such irreversible changes affected all areas of composites and can be explained as follows:

(i) Matrix: PLA was prone to undergo chemical (hydrolysis, cavitation) and mechanical (cracks, relaxations) damage under the combination of water and temperature.

In the presence of water, when temperature was close to the glass transition temperature, the high mobility of polymeric chains led to hydrolysis (Zhang et al., 2008). The loss in number-average molecular mass in Figure 8 indicates the chain scissions caused by hydrolysis. Besides, the mobility induced by both swelling (physi-crystallization) and chain scissions (chemi-crystallization) caused chains to reorganize in a more stable conformation. This crystallization of PLA is shown by DSC measurements. Indeed, PLA, which was completely amorphous after processing, reached a crystallinity of 6±2% after 144h in immersion at 50°C. Finally, swelling was prone to induce internal stresses which, in the presence of water, led to a mechanism called environmental stress cracking (Arnold, 1995). Figure 9 shows the presence of numerous cracks in PLA immersed at 50°C, suggesting that irreversible swelling (or density loss) in these samples was caused by swelling induced cracks.

(ii) Fibers: the hydrophilic behavior of plant fibers influences significantly the moisture sorption capacity of the composite. It is caused by their composition (polar groups prone to establish hydrogen bonds with water molecules) and their structure (porosity and high exchange surface) (Célino et al., 2014). As a result, when increasing the fiber content, the composites absorbs more water and the change of its properties is more likely to be irreversible. The dimensional stability of plant fibers is known to be highly dependent to moisture (Madsen et al., 2012). But their swelling is also highly anisotropic. For flax fibers, it can reach 25% in their transverse direction but rarely exceeds 0.2% in their longitudinal direction (Müssig, 2010). Hygroscopic strains are generally reversible but when they become substantial the micro-structure of cellulosic fibers undergo damage (Lee et al., 2010). Consequently, with such a significant macroscopic strain, it is very likely that flax fibers were damaged during the sorption and/or desorption process.

(iii) Interface fibers/matrix: the substantial swelling in the transverse direction of lignocellulosic fibers induced hoop stresses in the matrix, leading to plastic strain and micro-cracks (cf. Figure 10). The occurrence of micro-cracks has also been reported in several different nature of matrix (Azwa et al., 2013; Bismarck et al., 2002; Chow et al., 2007; Hu et al., 2010). The diffusion of water molecules was increased through the bulk matrix as well as along the fiber/matrix interfaces (Célino et al., 2014). Besides, water molecules are supposed to create hydrogen bonds with the fibers, which should have led to a reduction of fiber/matrix interactions (Dhakal et al., 2007). After drying, the shrinkage of fibers resulted in a gap between fibers and matrix visible in SEM pictures (cf. Figure 9). The increase in the irreversible swelling with fiber content in Table 2 is mainly the consequence of the void created by fibers shrinking during drying (le Duigou et al., 2012).

Besides, an extensive swelling of PLA led to shear stresses at the interfaces, due to the almost nonexistent swelling of fibers in their longitudinal direction. Such shear was likely to contribute to fiber/matrix decohesion (Le Duigou et al., 2015).

To sum up, the lack of dimensional stability of natural fibers to moisture is supposed to be the most decisive factors in the irreversibility of aging at the interface between fibers and matrix.

All these damage mechanisms are expected to lead to a drop of the mechanical properties. The likely consequences of these phenomena are assessed in the next part.

### 4.2. Reversibility/irreversibility of mechanical properties

The changes of physical properties during aging of materials indubitably affected their mechanical properties. Indeed, the increase in polymeric chains mobility caused by the diffusion of water between polymer chains induces a softening of polymeric materials (both fibers and matrix) called plasticizing (Weitsman, 2012). By comparing mechanical properties immediately before and after desiccation (cf. Table 1 and Table 3), it was possible to evaluate their reversibility, since plasticizing is the only reversible process occurring. Plasticizing turned out to be the main reason for the drop in DEM during immersion (except PLA-F30 at 35 and 50°C). Surprisingly, it turned out to be the only reason of the increase in damping during immersion, since the reversibility was complete (cf. Figure 7).

While reversible phenomena were quite dependent of the sole water content, the influence of irreversible mechanisms seemed to be more complex.

For PLA, the irreversible effects of aging were responsible for a slight increase in DEM with aging temperature. This observation was attributed to the increase in crystallinity measured by DSC and described previously. While aging did not affect the irreversibility of the ultimate tensile strength, it increased the irreversibility of the strain at break, probably due to stress relaxation induced by the forming process. However when aging temperature was close to the glass transition temperature, the irreversibility of the strain at break dropped. In this case, hydrolysis-caused damage prevailed (cf. Figure 8) which quickly led to a very high brittleness (immediate breaking). As a result, hydrolysis was essentially responsible for the irreversible loss of mechanical properties and its consequences occurred quite rapidly in these conditions.

For composites, the irreversible effects of aging should theoretically be enhanced compared to PLA due to the additional damaging mechanisms occurring in the presence of fibers. This hypothesis was mainly verified by the growing influence of irreversible mechanisms with fiber content. Damage induced by the presence of fibers became so important that irreversibility prevailed over reversibility in the drop of the DEM of PLA-F30 beyond 35°C.

Yet fibers did not always induce a more significant irreversibility. Indeed, composites immersed during 144h at 50°C exhibited an improvement in ductility while hydrolysis drastically deteriorated PLA. In addition, for longer aging times, the lifetime of materials was significantly extended by the presence of fibers which postponed the failure of PLA. Consequently, the damage engendered by the swelling of fibers in composites did not seem to be the main reason of the irreversible loss of properties (and thus of the failure mechanisms), but rather the larger amount of water (cf. Figure 2) which led to an advanced hydrolysis (cf. Figure 8). On the contrary, the presence of fibers significantly limited the propagation of cracks in PLA (Bourmaud et al., 2013).

## 5. CONCLUSIONS

This study exposed the effects of hydrothermal aging on PLA/flax biocomposites processed by extrusion and injection molding. It presented a methodology to distinguish the consequences of reversible and irreversible phenomena induced by hydrothermal aging of these biocomposites.

On the one hand, the reversible effects of aging were characterized by a limited swelling and a plasticizing of both PLA and flax fibers. On the other hand, the irreversible effects were characterized by a drop of PLA average molecular mass caused by hydrolysis which led to an increase in its crystallinity and the formation of cracks. In composites, it was accompanied by a substantial swelling of fibers leading to hoop stresses in the fibers vicinity and eventually decohesion from the PLA matrix.

Generally, the changes in mechanical properties were demonstrated to be primarily reversible. Damping turned out to be exclusively affected by the amount of water.

The irreversible effects of aging were strongly influenced by both the amount of fiber and the aging temperature. It became the primary cause for the drop of mechanical properties for the composite with the highest fiber content. Besides, the irreversible and reversible effects of aging did not always have the same effects on mechanical properties, i.e. while crystallizing induced a slight increase in elastic modulus, plasticizing induced a decrease. The use of the term "damage" implies some ambiguities, since such modification in the microstructure can lead to the increase of some mechanical properties (e.g. rigidity and ductility). Finally, even though the use of flax fibers in PLA led, for short-term aging, to a more significant irreversible decrease of several properties, it enabled to extend drastically the durability of such materials.

## ACKNOWLEDGMENTS

The authors would like to thank Frédéric Bonfils from CIRAD, UMR 1208 (IATE), Montpellier for his assistance with the size exclusion chromatography experiments.

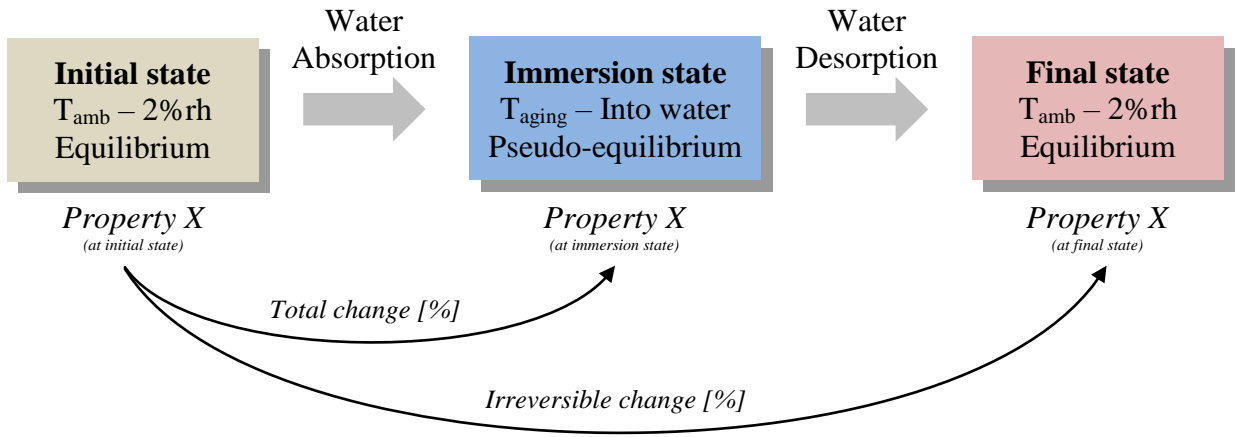

*Figure 1 – Schematic representation of the methodology to evaluate the irreversible changes in the characterized properties during hydrothermal aging*

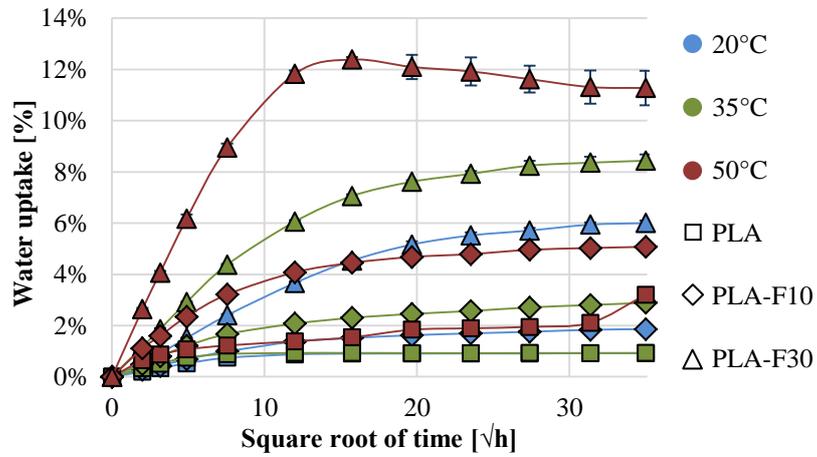

*Figure 2 – Water uptake of PLA and flax/PLA composites during immersion in water at different aging temperatures*

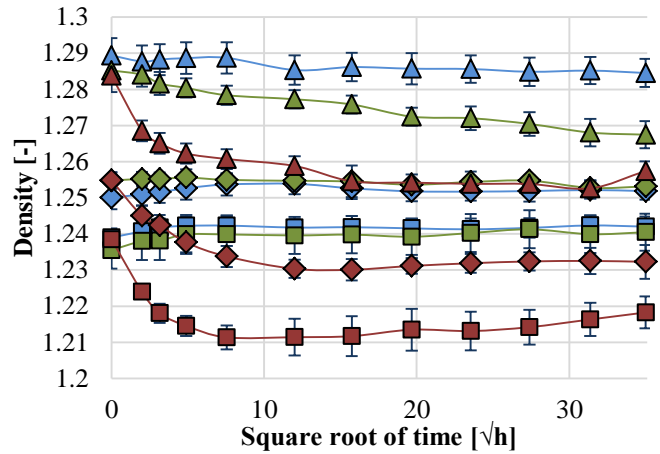

Figure 3 – Density of PLA and flax/PLA composites during immersion in water at different aging temperatures; the legend is the same as in Figure 2

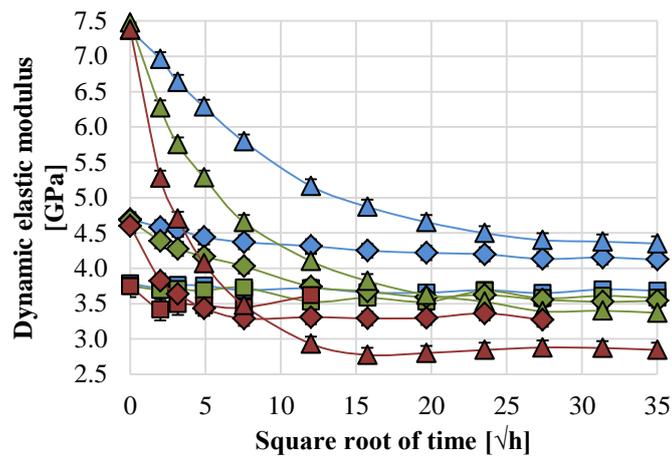

*Figure 4 – Dynamic elastic modulus of PLA and flax/PLA composites during immersion in water at different aging temperatures; the legend is the same as in Figure 2*

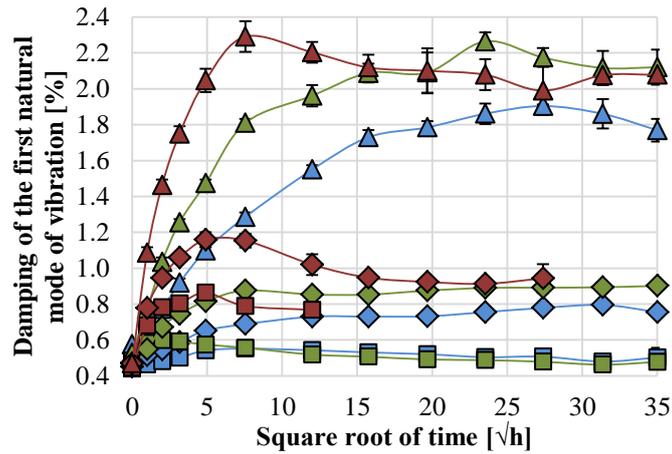

*Figure 5 – Damping (measured from the first fundamental natural mode of vibration) of PLA and flax/PLA composites during immersion in water at different aging temperatures; the legend is the same as in Figure 2*

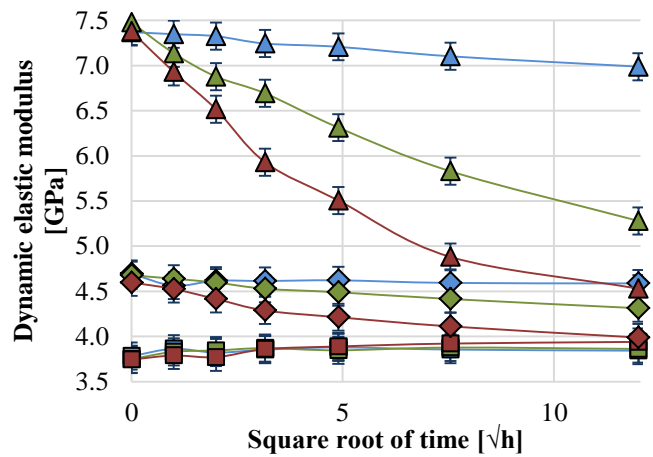

*Figure 6 – Dynamic elastic modulus of PLA and flax/PLA composites after immersion in water at different aging temperatures and complete desiccation; the legend is the same as in Figure 2*

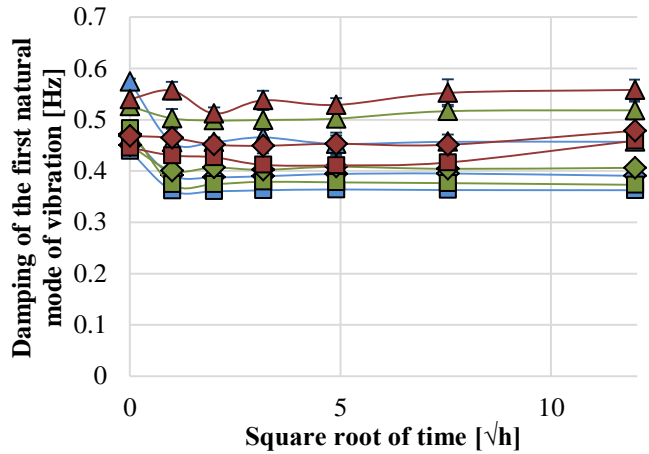

*Figure 7 – Damping (measured from the first fundamental natural mode of vibration) of PLA and flax/PLA composites after immersion in water at different aging temperatures and complete desiccation; the legend is the same as in Figure 2*

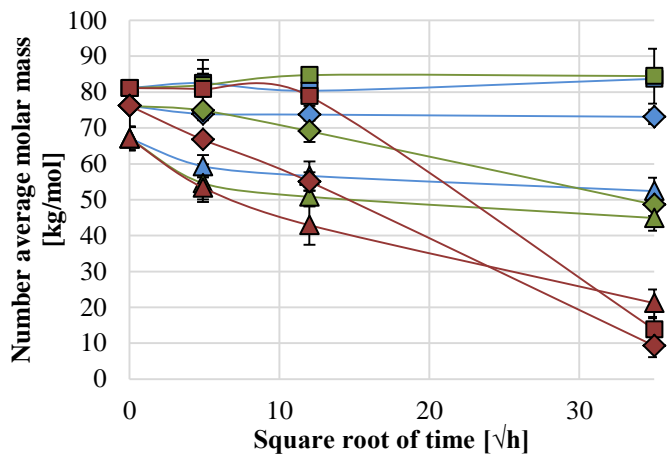

*Figure 8 – Number average molecular mass of PLA and flax/PLA composites during immersion in water at different aging temperatures; the legend is the same as in Figure 2*

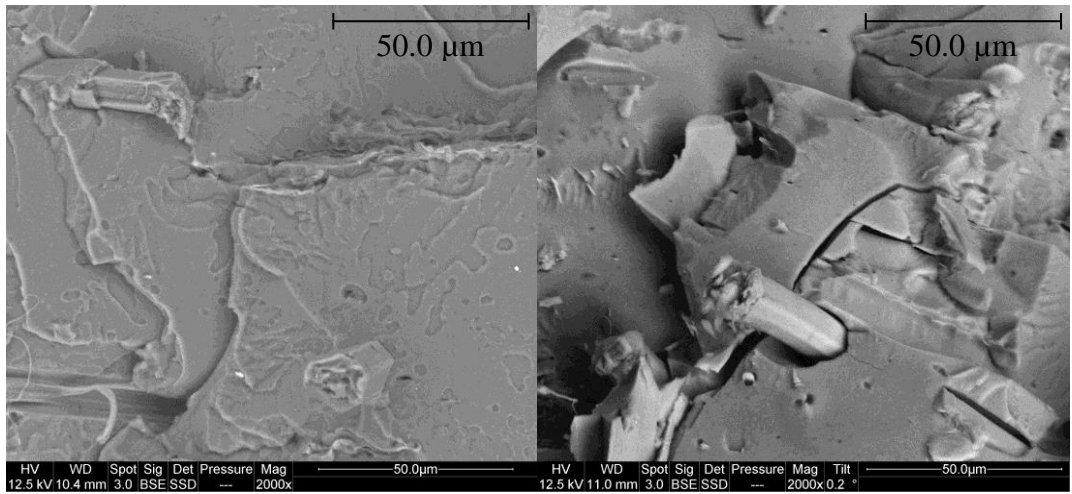

*Figure 9 – Freeze-fracturing surfaces of unaged (left) and aged (right) PLA/flax 10%wt composites.*

*The aged sample was immersed in water at 50°C during 1225h. (Magnification: ×2000)*

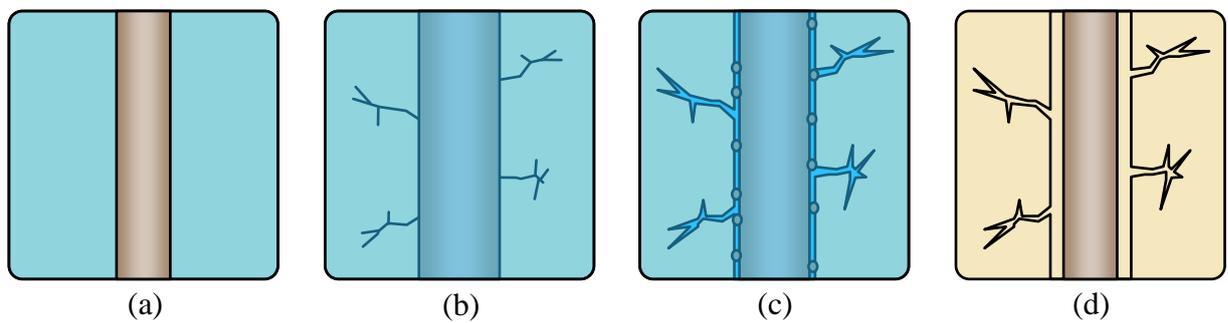

*Figure 10 – Damage mechanisms likely to occur in lignocellulosic composites during hydrothermal aging (adapted from (Azwa et al., 2013)) – (a) Water diffusion into the composite, (b) Swelling of fibers leading to matrix microcracks, (c) Leaching of water soluble substances from fibers, (d) Decohesion of fiber/matrix interface during drying*

| Aging Temperature | 20°C | | | 35°C | | | 50°C | | |
|---|---|---|---|---|---|---|---|---|---|
| Sample | PLA | PLA-F10 | PLA-F30 | PLA | PLA-F10 | PLA-F30 | PLA | PLA-F10 | PLA-F30 |
| Total change of UTS | -11% ± 2% | -20% ± 3% | -31% ± 2% | -29% ± 2% | -44% ± 3% | -53% ± 2% | - | - | - |
| Irreversible change of UTS | -1% ± 3% | -4% ± 3% | -6% ± 6% | -2% ± 3% | -13% ± 3% | -28% ± 6% | 3% ± 3% | -16% ± 3% | -41% ± 6% |
| Total change of SB | 96% ± 27% | 74% ± 29% | 77% ± 10% | 206% ± 83% | 229% ± 27% | 172% ± 30% | - | - | - |
| Irreversible change of SB | 52% ± 22% | 12% ± 3% | 14% ± 15% | 36% ± 36% | 38% ± 12% | 69% ± 18% | -51% ± 2% | 38% ± 2% | 78% ± 1% |

*Table 1 – Irreversible and total changes of ultimate tensile strength (UTS) and strain at break (SB) of PLA and flax/PLA composites after 144h of aging*

| Aging Temperature | 20°C | | | 35°C | | | 50°C | | |
|---|---|---|---|---|---|---|---|---|---|
| Sample | PLA | PLA-F10 | PLA-F30 | PLA | PLA-F10 | PLA-F30 | PLA | PLA-F10 | PLA-F30 |
| Irreversible change of mass | 0.00% ± 0.01% | 0.00% ± 0.01% | 0.01% ± 0.01% | 0.00% ± 0.01% | 0.01% ± 0.01% | 0.05% ± 0.04% | 0.04% ± 0.01% | 0.04% ± 0.01% | 0.04% ± 0.04% |
| Irreversible change of volume | -0.42% ± 1.40% | 0.10% ± 0.54% | 0.25% ± 0.26% | 0.52% ± 1.00% | 1.46% ± 0.42% | 4.36% ± 0.54% | 2.67% ± 1.79% | 5.49% ± 0.63% | 10.06% ± 0.27% |
| Irreversible change of density | 0.01% ± 0.28% | -0.05% ± 0.10% | -0.26% ± 0.22% | 0.16% ± 0.47% | -1.20% ± 0.14% | -4.01% ± 0.31% | -2.42% ± 0.37% | -5.06% ± 0.22% | -9.42% ± 0.12% |

*Table 2 – Irreversible changes of morphological properties (mass, volume and density changes) of PLA and flax/PLA composites after 144h of aging*

| Aging Temperature | 20°C | | | 35°C | | | 50°C | | |
|---|---|---|---|---|---|---|---|---|---|
| Sample | PLA | PLA-F10 | PLA-F30 | PLA | PLA-F10 | PLA-F30 | PLA | PLA-F10 | PLA-F30 |
| Total change | -1,7% | -8,1% | -26,0% | -5,9% | -20,0% | -45,1% | -3,4% | -28,1% | -60,3% |
| Irreversibility | 1,6% | -2,3% | -9,6% | 3,0% | -7,8% | -28,4% | 5,2% | -13,3% | -38,6% |
| Reversibility | -3,3% | -5,9% | -16,4% | -8,9% | -12,2% | -16,8% | -8,7% | -14,8% | -21,7% |

*Table 3 – Reversible and irreversible changes of dynamic elastic modulus of PLA and flax/PLA composites after 144h of aging*